\documentclass[iop]{emulateapj}

\usepackage{graphicx}
\usepackage{stmaryrd}
\usepackage{multirow}

\usepackage{natbib}
\newcommand{\simi}{\ensuremath{\sim}}

\newcommand{\hd}{HD~189733}
\newcommand{\hdsp}{HD~189733 }
\newcommand{\HST}{\textit{HST}}
\newcommand{\HSTsp}{\textit{HST} }
\newcommand{\m}{$\mu$}
\newcommand{\eg}{\textit{e.g.} }
\newcommand{\ie}{\textit{i.e.} }
\newcommand{\modif}{}
\newcommand{\mod}{}
\newcommand{\modi}{}
\newcommand{\modd}{}

\shorttitle{The eclipse of HD~189733\MakeLowercase{$\rm b$}}
\shortauthors{Crouzet et al.}

\begin{document}

\title{Water vapor in the spectrum of the extrasolar planet HD~189733\MakeLowercase{b}: 2. The eclipse}

\author{Nicolas Crouzet \altaffilmark{1,2}}
\author{Peter R. McCullough \altaffilmark{2,3}}
\author{Drake Deming \altaffilmark{4,5}}
\author{Nikku Madhusudhan \altaffilmark{6}}

\affil{\altaffilmark{1} Dunlap Institute for Astronomy \& Astrophysics, University of Toronto, 50 St. George Street, Toronto, Ontario, Canada M5S 3H4}
\affil{\altaffilmark{2} Space Telescope Science Institute, Baltimore, MD 21218, USA}
\affil{\altaffilmark{3} Department of Physics and Astronomy, Johns Hopkins University, 3400 North Charles Street, Baltimore, MD 21218, USA}
\affil{\altaffilmark{4} Department of Astronomy, University of Maryland, College Park, MD 20742, USA}
\affil{\altaffilmark{5} NASA Astrobiology Institute's Virtual Planetary Laboratory, USA}
\affil{\altaffilmark{6} Institute of Astronomy, University of Cambridge, Madingley Road, Cambridge, CB3 0HA, UK}

\email{crouzet@dunlap.utoronto.ca}

\begin{abstract}

Spectroscopic observations of exoplanets are crucial to infer the composition and properties of their atmospheres. \hd b is one of the most extensively studied exoplanets and is a corner stone for hot Jupiter models. In this paper, we report the day-side emission spectrum of \hd b in the wavelength range 1.1 to 1.7~\m m obtained with the \textit{Hubble Space Telescope} Wide Field Camera~3 in spatial scan mode. The quality of the data is such that even a straightforward analysis yields a high precision Poisson noise limited spectrum: the median 1-$\sigma$ uncertainty is 57 ppm per 0.02~\m m bin. We also build a white-light curve correcting for systematic effects and derive an absolute eclipse depth of $96\pm39$~ppm. The resulting spectrum shows \modi{marginal evidence for water vapor absorption, but can also be well explained by a blackbody spectrum. However, the combination of these WFC3 data with previous \textit{Spitzer} photometric observations is best explained by a day-side atmosphere of \hd b with no thermal inversion and a nearly solar or sub-solar H$_2$O abundance in a cloud-free atmosphere. Alternatively, this apparent sub-solar abundance may be the result of clouds or hazes which future studies need to investigate.} 

\end{abstract}


\section{Introduction}

Spectroscopic observations of exoplanets are crucial to infer the composition and properties of their atmospheres. The eclipse, when the planet passes behind the star, probes the day-side emission spectrum of the planet's atmosphere. The expected signal for molecular signatures is a few 100 ppm; this extreme precision is currently achievable only for planets orbiting bright stars. As a result, only a handful of exoplanets have been characterized spectroscopically. With a host star of magnitude J=6.07, \hd b is one of the most extensively studied exoplanets along with HD 209458b. \hd b is a 1.144 M$_{jup}$, 1.138 R$_{jup}$ gas giant planet orbiting an active K dwarf in 2.22 days \citep{Bouchy2005}. Spectra and phase curves obtained at several wavelengths delivered quantities of information about its atmosphere. \citet{Grillmair2007,Grillmair2008} obtained spectra in emission with \textit{Spitzer} revealing strong water vapor absorption and the presence of an extra-absorber in the day-side upper atmosphere. A hot spot was detected eastward of the substellar point revealing an equatorial super-rotating jet \citep{Knutson2007}, as anticipated by atmospheric circulation models of hot Jupiters \citep{Showman2002}. Brightness temperature measurements constrained the pressure-temperature profile of both sides of the atmosphere and the day-night heat redistribution efficiency \citep{Knutson2012,Desert2011}. In addition, signatures of non-equilibrium chemistry were reported \citep{Knutson2012,Swain2010}. Several molecules were identified in the planet's transmission and emission spectra with \textit{Hubble Space Telescope (\HST)} NICMOS \citep{Swain2008,Swain2009,Swain2014}. However, subsequent studies of NICMOS data concluded that instrumental systematics may contribute significantly to the observed spectra \citep{Gibson2011b,Crouzet2012}. \modd{More recently, CO and H$_2$O were detected in the day-side of the planet's atmosphere using ground-based high-resolution spectroscopy with Keck II NIRSPEC and VLT CRIRES \citep{Rodler2013,deKok2013,Birkby2013}}. A new era began for exoplanet spectroscopy with the installation of the Wide Field Camera~3 (WFC3) onboard \HSTsp in 2009. \citet{Gibson2012} observed two transits of \hd b with WFC3. Unfortunately, the central part of the spectrum saturated the detector; only the edges were used to derive the planetary radius in two wavelength ranges. The recently implemented spatial scanning mode \citep{McCullough2012} now allows WFC3 to observe \hdsp in much better conditions: this mode is designed to obtain high sensitivity on bright stars for exoplanet spectroscopy by reducing overheads and avoiding detector saturation. Observations of \hdsp with WFC3 in spatial scanning mode (HST program 12881, PI: McCullough) were obtained during a planetary transit \citep{McCullough2014} and during a planetary eclipse (this paper).


Although clear atmosphere models are often used to interpret exoplanetary spectra, recent results suggest that clouds and hazes may play an important role in the atmosphere of exoplanets. Atmospheric clouds would result in an increased opacity which may obscure individual molecular spectral features. As examples, spectra of the hot Jupiters HD 209458b and XO-1b with \HSTsp WFC3 in spatial scanning mode showed a water vapor absorption feature at 1.4 \m m with an amplitude much smaller than expected from clear atmosphere models \citep{Deming2013}. The extreme case of an absence of molecular features would indicate an atmosphere completely dominated by opaque clouds. Using 2-channel near-infrared photometry with NICMOS during transits of \hd b, \citet{Sing2009} found results consistent with Rayleigh scattering from haze. No water vapor absorption was detected. Similarly, \citet{Gibson2012} suggested that haze may dominate its near-infrared transmission spectrum. Higher layers of the atmosphere of \hd b have also been probed. High altitude haze was inferred by a UV transmission spectrum dominated by Rayleigh scattering \citep{Lecavelier2008,Sing2011,Huitson2012}, which would explain the narrowness of alkali features and the increased planetary radius in the UV compared to the infrared. After re-analyzing \hdsp data over a wide wavelength range, \citet{Pont2013} proposed a general explanation for the atmosphere of \hd b at its terminator: Rayleigh scattering by clouds dominates in the UV, and settling dust and a cloud deck yield a featureless spectrum in the near infrared and in the infrared. A marginal detection of a decreasing geometric albedo in the visible may also indicate optically thick reflective clouds on the day-side of the planet, although with a low albedo overall \citep{Evans2013}. Beyond the case of hot Jupiters, examples of featureless transmission spectra are GJ1214b \citep{Berta2012,Kreidberg2014}, GJ436b \citep{Knutson2014a}, and HD 97658b \citep{Knutson2014b}, \modif{although for low-mass planets, flat transmission spectra could alternatively be explained by high mean molecular weight atmospheres. In all cases, the inferences of hazes/clouds are motivated by the non-detection of spectral features of expected molecules, and a blueward rise in the optical part of the transmission spectrum. However, observational constraints on the chemical composition of the haze/cloud material are non-existent}. In this context, measuring the transit and eclipse spectra of \hd b in the near infrared, where molecular features are expected, is crucial. The observations presented in this paper and in \citet{McCullough2014} bring new constraints on \modif{the cloud hypothesis for} the atmosphere of \hd b. 

For tidally locked exoplanets such as \hd b, one might expect different atmospheres on the day side, the nightside, and along the terminator. However, very strong equatorial winds would tend to homogenize both sides by increasing the heat redistribution and mixing the composition. In the case of \hd b, the heat redistribution was found to be relatively small \citep{Grillmair2008,Knutson2012}. Measuring the transit and eclipse spectra of \hd b in the same wavelength range with the high precision of WFC3 spatial scans offers a unique opportunity to consistently probe these two regions of the atmosphere.

We present the observations in section \ref{sec: Observations}, the data reduction in section \ref{sec: Data reduction}, the resulting spectrum in section \ref{sec: Results}. We perform a whitelight analysis in section \ref{sec: whitelight analysis}. Results are then discussed in section \ref{sec: Discussion}, followed by a summary in section \ref{sec: Summary}.


\section{Observations}
\label{sec: Observations}

\begin{figure}[htbp]
   \centering
   \includegraphics[width=8cm]{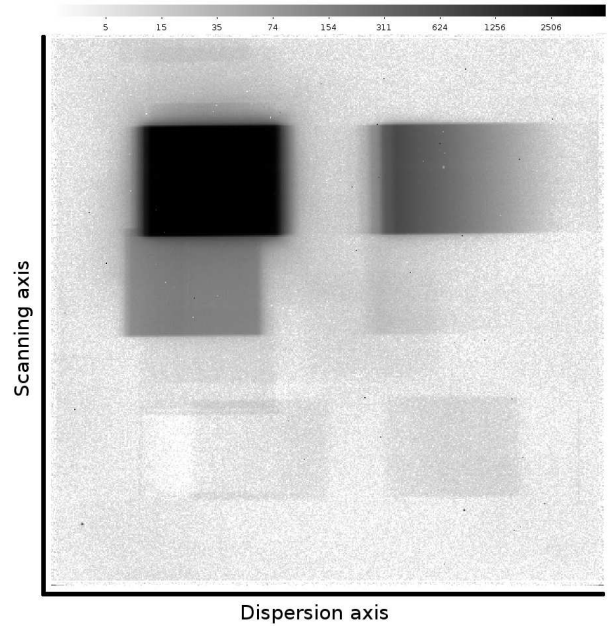}
      \caption{Example of image acquired with WFC3 using the spatial scanning mode. The wavelengths are spread horizontally whereas the spatial scan occurs vertically. The intensity is in log(ADU) s$^{-1}$ as indicated at the top. The first order spectrum of \hd A, used is in this work, is located in the upper left quadrant. Also visible are the second order spectrum of \hd A in the upper right quadrant, the first and second order spectrum of the companion star \hd B below and slightly overlapping with those of \hd A, and the spectra of fainter stars.}
   \label{fig: spatial scanning}
\end{figure}

We used \HSTsp WFC3 with the newly implemented spatial scanning mode, developed in part to enable observations such as these \citep{McCullough2012}. In this mode, a controlled scan is applied to the telescope during the exposure in a direction perpendicular to the wavelength dispersion direction (Figure \ref{fig: spatial scanning}). This technique is particularly efficient for bright stars such as \hdsp \citep[see][for more details]{McCullough2014}. Several programs have already benefited from this technique \citep{Deming2013,Wakeford2013,Kreidberg2014,Knutson2014a,Knutson2014b}.

One eclipse of \hd b was observed on June 24, 2013. The observations are divided into five \HSTsp orbits, the planetary eclipse occurring during the fourth orbit. In total, 159 exposures of 5.97 s each were acquired, corresponding to 32 exposures per orbit (except for the first orbit in which the first image is a direct image). We used the G141 grism covering a spectral range from 1.1 to 1.7 \m m, the 512$\times$512 pixel subarray, and a scan rate of 2 arcsec s$^{-1}$.  The spectral trace is spread over 111 detector rows by the spatial scan and 150 columns by the dispersive element. The resulting spectral resolution is $R=\lambda /{\Delta\lambda} = 130$ \citep{Dressel2014}. The detector is read out in MULTIACCUM mode with the RAPID sample sequence and NSAMP = 7 (each image is composed of 7 successive readouts after the initial read). The first and second order spectra of \hdsp are visible in the images, as well as the spectra of the companion star \hd B. The latter slightly overlap with the former, but our data reduction method eliminates this overlap. The spatial scans occur alternatively in two directions, which we will designate as ``forward" and ``reverse" in this paper. The observation parameters are summarized in Table \ref{tab: observations parameters}.

\begin{table}
\begin{center}
\caption{Summary of \HSTsp WFC3 observations \label{tbl-1}}
\label{tab: observations parameters}
\begin{tabular}{lc}
\tableline
\tableline
 \HSTsp Program (P.I.) & 12881 (McCullough) \\
 Number of \HSTsp orbits & 5 \\
 Number of scans per orbit  &  16 Forward, 16 Reverse \\
 Duration of scan (s) & 5.97 \\
 Scan rate (arcsec s$^{-1}$)[pixels s$^{-1}$] & (2.00)[16.5] \\
 Peak signal on detector (e$^{-}$px$^{-1}$) & $4.0\times 10^4$ \\
 Grism ($\lambda$) & G141 ($1.1-1.7 \mu$m) \\
 Detector subarray size (pixels) & 512x512 \\
 Sample sequence & RAPID \\
 Samples per scan & 8 \\
 Start of first scan (HJD)  & 2456467.855665  \\
 Corresponding planetary orbital phase & 0.40717 \\
 Start of last scan (HJD)  & 2456468.144658  \\ 
 Corresponding planetary orbital phase &  0.53743 \\
\tableline
\end{tabular}
\end{center}
\vspace{-2mm}
Notes.
Forward and reverse scans were interleaved.
The planetary orbital phase is defined to be 0.5 at mid-eclipse.
\end{table}

\section{Data reduction}
\label{sec: Data reduction}

\subsection{Spectrum extraction}

We intentionally keep the data reduction as simple as possible in order to emphasis the data quality obtained with \HSTsp WFC3 in spatial scanning mode. This is in stark contrast with previous similar observations such as those obtained with \HSTsp NICMOS \citep[\textit{e.g.}][]{Crouzet2012}, in which complex data reduction methods were necessary. The improvement of the data quality of WFC3 compared to NICMOS is already evident after a first look at the images: variations of the spectrum position on the detector are nearly invisible to the eye and the individual pixel values are much more stable.

\begin{figure}[htbp]
   \centering
   \includegraphics[width=8cm]{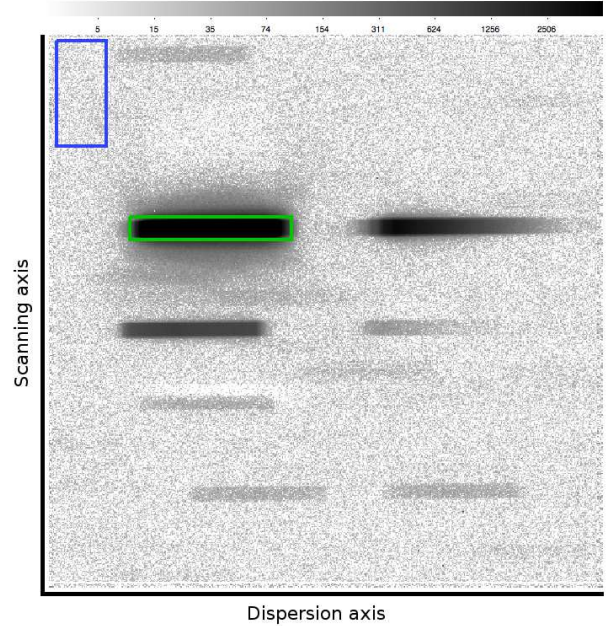}
      \caption{Example of a differential image obtained after subtraction of two consecutive readouts of the image shown in Figure \ref{fig: spatial scanning}. The extension of the various spectral traces along the scanning direction is now reduced, as the full scan is now split between 7 differential images (see text). The intensity is in log(ADU) s$^{-1}$ as indicated at the top. The stellar and sky regions are shown in green and blue respectively.}
   \label{fig: differential images}
\end{figure}

We start with the Intermediate MultiAccum images (ima.fits), which are corrected for dark current, non-linearity, and other calibrations by the CALWFC3 pipeline at STScI. We separate the images according to the scanning direction, \modif{resulting in two sets of data. Then, for each image, we build seven differential images from the eight readouts. Indeed, the WFC3 pixels are read non-destructively while the signal is building up and while the vertical scanning occurs, and eight intermediate readouts are recorded during each exposure. The first readout occurs quickly after the exposure starts and is known as the zeroth read, whereas the last readout corresponds to the full scan image. The signal recorded between two consecutive readouts is retrieved by subtracting them, \ie $D_i = R_{i} - R_{i-1}$, where $R_i$ are the readouts, $D_i$ are the ``differential images", and $i = 1,...,7$. Figure \ref{fig: differential images} shows an example of a differential image. Its size along the scan direction is now reduced as it corresponds to 1/7th of the full scan. Considering all the observations, this leads to seven sets of differential images for each scan direction, \ie fourteen sets in total. Within each set, the spectral trace is always at the same position on the detector. We build a 1-D spectrum for each set independently. This method eliminates the overlap of the companion star with the target, minimizes the sky background contribution, and allows consistency checks between the spectra obtained from the fourteen sets.}

We calibrate each differential image \modi{($D_i$)} by the F139M flat-field. We consider the pixel response is not wavelength dependent at the level required to extract the planetary spectrum, and we simply divide each image by this two-dimensional flat-field. Further tests using a wavelength dependent flat-field or no flat-field calibration do not result in significant differences in the final spectrum. Furthermore, flat-fielding errors should in principle be largely removed by dividing the in- by the out-of-eclipse spectra. 

We search for bad pixels and cosmic rays located in the spectral trace region on a row-by-row basis. We compute a median-smoothed version of the row using a 10-pixel kernel. We calculate the median of the absolute deviation of the row with respect to its smoothed version and identify pixels deviating by more than 15 times this median. This yields typically 2 to 3 deviant pixels over the spectral region in each differential image, which are in fact already apparent by eye. We replace them by the value of the median-smoothed row at this location. We find good agreement between the bad pixels identified by our method and those flagged by the CALWFC3 pipeline as bad pixels in the data quality arrays, \textit{i.e.} with a flag value of 4: ``Bad detector pixel'' \citep{Rajan2010} in extension 3 of the ima.fits files.

As commonly done with \HST, we do not use the first orbit in which the telescope is known to settle to its new thermal environment resulting in an unstable behavior. As a result, the ``hook" observed in the whitelight curve at the beginning of each orbit, a WFC3 feature already reported by \citet{Berta2012} and \citet{Deming2013}, is much stronger in the first orbit. All images from the four remaining orbits are used in the analysis.

We define rectangular regions containing the stellar signal and the sky background. A stellar region is defined for each set of differential images. The regions are 154 columns wide and 20 rows tall for forward scans \modif{(22 rows tall for reverse scans). The reverse direction yields slightly taller scans due to the longer exposure time \citep{McCullough2012}} . The sky region is common to all images and is 96 columns wide, 166 rows tall, and located in the upper left corner. 

\begin{figure}[htbp]
   \centering
   \includegraphics[width=8cm]{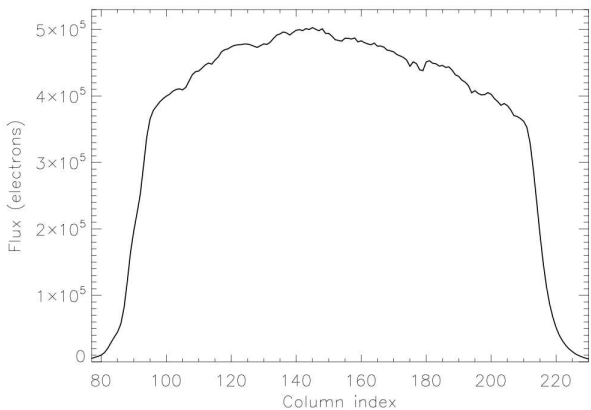}
      \caption{Example of spectrum obtained from a differential image \modi{($D_i$)}.}
   \label{fig: raw spectrum}
\end{figure}

Spectra are built in a very straightforward manner. The sky background is calculated on the sky region using the SKY procedure in IDL and subtracted to the image. The sky values range from $0.96\times10^{-4}$ to $1.71\times10^{-4}$ relatively to the maximum flux per pixel in the spectral trace of \hd. Then, we sum each column in the stellar region individually; this creates a one-dimensional spectrum for each differential image \modi{$D_i$} (figure \ref{fig: raw spectrum}). For each set of differential images, we average these spectra for each orbit, and average the three out-of-eclipse orbits together (only one orbit is in-eclipse). Then we divide the in-eclipse spectrum by the out-of-eclipse spectrum to extract the planetary spectrum. We build 14 planetary spectra from the 14 sets of differential images, which we average together. For all these calculations, we use the most simple possible average (function MEAN in IDL), with \eg no calls for outliers nor medians\footnote{However, recall that we identified and replaced outlier pixel values in the 2-D images.}. 

\subsection{Wavelength calibration}

\begin{figure}[htbp]
   \centering
   \includegraphics[width=8cm]{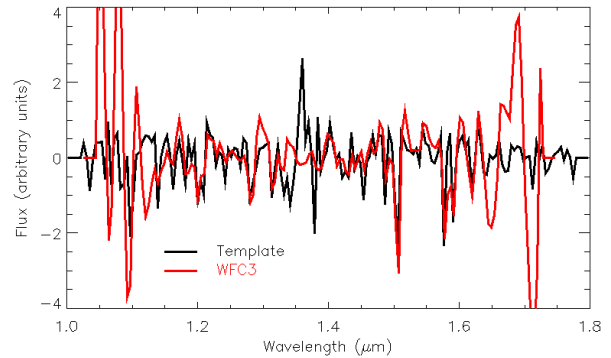}
      \caption{Wavelength calibration. Black: library spectrum of the K1V star 107 Psc. Red: stellar spectrum extracted from the WFC3 data. The large spikes in the WFC3 data at the ends of the bandpass are due to the low response of the grism at the edges.}
   \label{fig: wavelength solution}
\end{figure}

To calibrate the detector columns in wavelength, we match our stellar spectrum to a known spectrum of similar spectral type. In keeping with our objective of simplicity, the G141 grism wavelength dispersion can be approximated by a linear function.
First, we calibrate the out-of-eclipse spectrum by the known G141 grism response. This requires an initial estimate of the wavelength solution, obtained by matching the sharp edges of the grism response to that of our spectrum. Second, we compare a spectrum of the K1V star 107 Psc\footnote{http://irtfweb.ifa.hawaii.edu/$\scriptstyle\sim$spex/IRTF\_Spectral\_Library/\\index.html} \citep{Rayner2009} to our data. We remove the low frequency variations by subtracting a smoothed version from each of these spectra. Finally, we match them in the wavelength range $1.1-1.7 \mu$m using their common spectral lines (Figure \ref{fig: wavelength solution}). We derive the relation $\lambda(x)=a\times(x-x_0) + b$, where $x$ is the detector column index, $x_0=77$, $a=0.0047$ \m m px$^{-1}$, and $b=1.027$ \m m.

\subsection{Spectrum shifts}

We investigate the variable shifts in the $x$ direction (along the rows), \modif{which may be present in WFC3 spatial scan data \citep{Deming2013}. To calculate these, we re-create full scan images by summing the calibrated forward differences, and build a template spectrum for each scan direction from the 40th such image. We then calculate the shift of each row with respect to the template spectrum. The row is shifted by steps of 1/100 pixel up to $\pm 1$ pixel} using linear interpolation; at each step, the shifted spectrum is divided by the template, normalized by its median, and the standard deviation of this ratio is calculated. The smallest standard deviation defines the best matching shift. The original row is then replaced by the corresponding interpolated row in each forward difference image, and a shift-corrected spectrum is built. Overall, we find a shift mean absolute deviation of 0.014 and 0.013 pixel for the forward and reverse scan directions respectively. The final column to column planetary spectra are nearly identical with or without this correction. As this yields negligible benefit and may instead introduce spurious effects, we choose to remain as close to the data as possible and do not include this correction in our final spectrum. In contrast with NICMOS where the variations of the position of the spectral trace on the detector were a strong limiting factor \citep{Crouzet2012}, such variations are here barely detected as they remain at the 1/100 pixel level, and they do not affect the planetary spectrum.

\subsection{Uncertainties}
\label{sec: errors}

We estimate the uncertainties of the exoplanetary spectrum \modif{from the standard deviation of the lightcurve of each spectral channel (a detector column or group of binned columns corresponding to a given wavelength). These lightcurves are first corrected from whitelight flux variations. To this end, we extract a spectrum for each calibrated full scan image excluding 20 columns at each edge of the spectral trace, and scale its amplitude to that of the template. The best scaling factor is found using the downhill simplex minimization procedure (AMOEBA). We build the whitelight corrected channel lightcurves from these amplitude corrected spectra. The uncertainty $\sigma_{c,d}$ for a given channel $c$ and a given scan direction $d$ is:}

\begin{equation}
\sigma_{c,d} = \sqrt{\frac{\sigma_{out,d}^2}{m_{out,d}^2 N_{out,d}} + \frac{\sigma_{in,d}^2}{m_{in,d}^2 N_{in,d}}}
\end{equation}
\modif{where $\sigma_{in,d}$, $\sigma_{out,d}$ are the standard deviations of the in-eclipse and out-of-eclipse parts of the channel $c$ lightcurve respectively, $m_{in,d}$, $m_{out,d}$ their respective mean, and $N_{in,d}$, $N_{out,d}$ their respective number of points ($N_{in,d}=16$, $N_{out,d}=48$). We find a median uncertainty of 160 ppm and 146 ppm for the forward ($d=f$) and reverse ($d=r$)} scan directions respectively, for an expected minimum Poisson noise of 162 and 153 ppm. Our final spectrum is the average of the forward and reverse scan direction spectra, thus our final uncertainties are: 

\begin{equation}
\sigma_c = \frac{1}{2}\sqrt{\sigma_{c,f}^2+ \sigma_{c,r}^2}
\end{equation}
The final median uncertainty is 110 ppm compared to a Poisson noise estimate of 111 ppm. Because the planet passes behind the star during the eclipse, the effect of star spots on the \modif{emission} spectrum can be neglected.

\section{Results}
\label{sec: Results}

\begin{figure}[htbp]
   \centering
   \includegraphics[width=8.5cm]{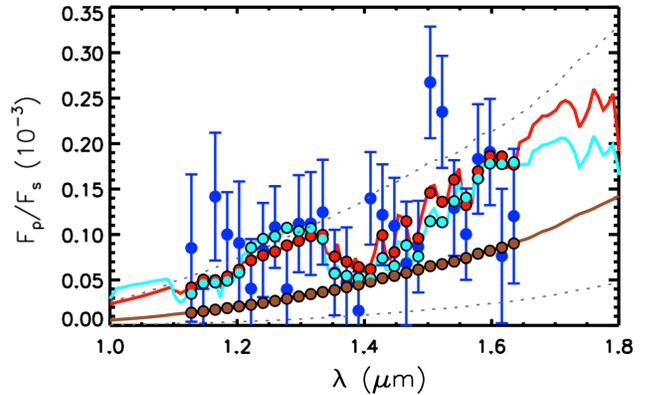}
      \caption{Day-side emission spectrum of HD 189733b obtained with WFC3 (blue), and theoretical model spectra for a clear planetary atmosphere of solar composition (cyan), and close to solar composition (red), with their averages at the WFC3 wavelength bins (filled circles). \modif{The best-fit blackbody spectrum for the WFC3 and {\it Spitzer} data together (see Section \ref{sec: Discussion}), at 1295~K, is also shown (brown)}. The dashed lines are blackbody spectra at 1100~K (bottom) and 1500~K (top). The vertical axis is the ratio of planetary to stellar flux.}
   \label{fig: final spectrum}
\end{figure}

The planetary emission spectrum is binned over 4 columns using a boxcar average, resulting in independent 0.02 \m m bins. \modif{This yields a high precision Poisson noise limited spectrum: the median uncertainty is 57 ppm. This spectrum is reported in Table~\ref{tab: final spectrum} columns 1 to 4, with an average depth arbitrarily set to zero, and is shown in Figure \ref{fig: final spectrum}. The absolute eclipse depth is calculated from the whitelight analysis in Section \ref{sec: whitelight analysis}. The spectra obtained from the 2 scan directions and for the 14 sets of differential images are consistent within their uncertainty and do not show any systematic differences.}

An independent analysis of these data was conducted using the method prescribed in \citet{Deming2013}. Both planetary spectra are largely consistent, well within their 1-$\sigma$ uncertainties. This second spectrum as well as the difference between both are reported in Table \ref{tab: final spectrum}, columns 5 to 7. The smaller uncertainties of this second spectrum are due to the convolution with a Gaussian kernel of Full Width Half Maximum of 4 columns used to balance the effect of under-sampling \citep[see][]{Deming2013}, which is equivalent to an effective binning over \simi6 columns, whereas the spectrum in columns 1 to 3 is binned over 4 columns.

\begin{table*}
\begin{center}
\caption{Day-side emission spectrum of \hd b}
\begin{tabular}{rrrrrrr}

\hline
\hline
 $\lambda$  & $\Delta R_p^2/R_s^2$ & $\sigma$ & Column &  $\Delta R_p^2/R_s^2$ & $\sigma$ & $\Delta\Delta$ \\
 ($\mu$m)  & (ppm) & (ppm) & & (ppm) & (ppm) & (ppm) \\
\hline

1.1279  &   -21  &    81   &   98.5  &   -96  &    47  &    75  \\
1.1467  &  -127  &    61   &  102.5  &   -18  &    50  &  -109  \\
1.1655  &    36  &    70   &  106.5  &    28  &    45  &     8  \\
1.1843  &    -6  &    47   &  110.5  &    -3  &    44  &    -3  \\
1.2031  &   -15  &    68   &  114.5  &    -7  &    43  &    -9  \\
1.2218  &   -65  &    55   &  118.5  &   -45  &    50  &   -20  \\
1.2406  &   -23  &    52   &  122.5  &   -32  &    42  &     9  \\
1.2594  &     2  &    45   &  126.5  &     3  &    42  &    -1  \\
1.2782  &   -66  &    50   &  130.5  &   -16  &    42  &   -51  \\
1.2969  &     6  &    53   &  134.5  &    53  &    41  &   -47  \\
1.3157  &     6  &    57   &  138.5  &    31  &    41  &   -24  \\
1.3345  &    19  &    58   &  142.5  &    12  &    41  &     7  \\
1.3533  &   -48  &    47   &  146.5  &   -21  &    41  &   -27  \\
1.3721  &   -59  &    61   &  150.5  &   -27  &    41  &   -32  \\
1.3908  &   -90  &    43   &  154.5  &   -40  &    42  &   -50  \\
1.4096  &    34  &    51   &  158.5  &     8  &    42  &    26  \\
1.4284  &    16  &    55   &  162.5  &    -2  &    42  &    18  \\
1.4472  &     4  &    53   &  166.5  &   -29  &    42  &    33  \\
1.4660  &   -38  &    68   &  170.5  &   -64  &    43  &    26  \\
1.4848  &   -19  &    50   &  174.5  &     9  &    43  &   -29  \\
1.5035  &   161  &    61   &  178.5  &   132  &    43  &    29  \\
1.5223  &   129  &    62   &  182.5  &   115  &    43  &    14  \\
1.5411  &    23  &    53   &  186.5  &    47  &    44  &   -23  \\
1.5599  &    -6  &    50   &  190.5  &     3  &    45  &    -8  \\
1.5786  &    77  &    60   &  194.5  &    39  &    45  &    38  \\
1.5974  &    85  &    58   &  198.5  &   -16  &    46  &   101  \\
1.6162  &   -30  &    74   &  202.5  &   -64  &    46  &    35  \\
1.6350  &    14  &    74   &  206.5  &     --  &     --  &     --  \\

\hline  
\label{tab: final spectrum}
\end{tabular}
\end{center}
\vspace{-5mm}
Notes. Units are as indicated; parts per million is abbreviated ppm.  
The tabulated uncertainties apply to the differential eclipse depths; an
 additional uncertainty applies to the overall depth (see text). 
The first three columns refer to the analysis of N. C.; columns 5 and 6 refer to the analysis of D. D.;
the last column contains the difference of the differential spectra, column 2 minus column 5.
\end{table*}

\section{Whitelight analysis}
\label{sec: whitelight analysis}

\begin{figure}[htbp]
   \centering
   \includegraphics[width=4.1cm]{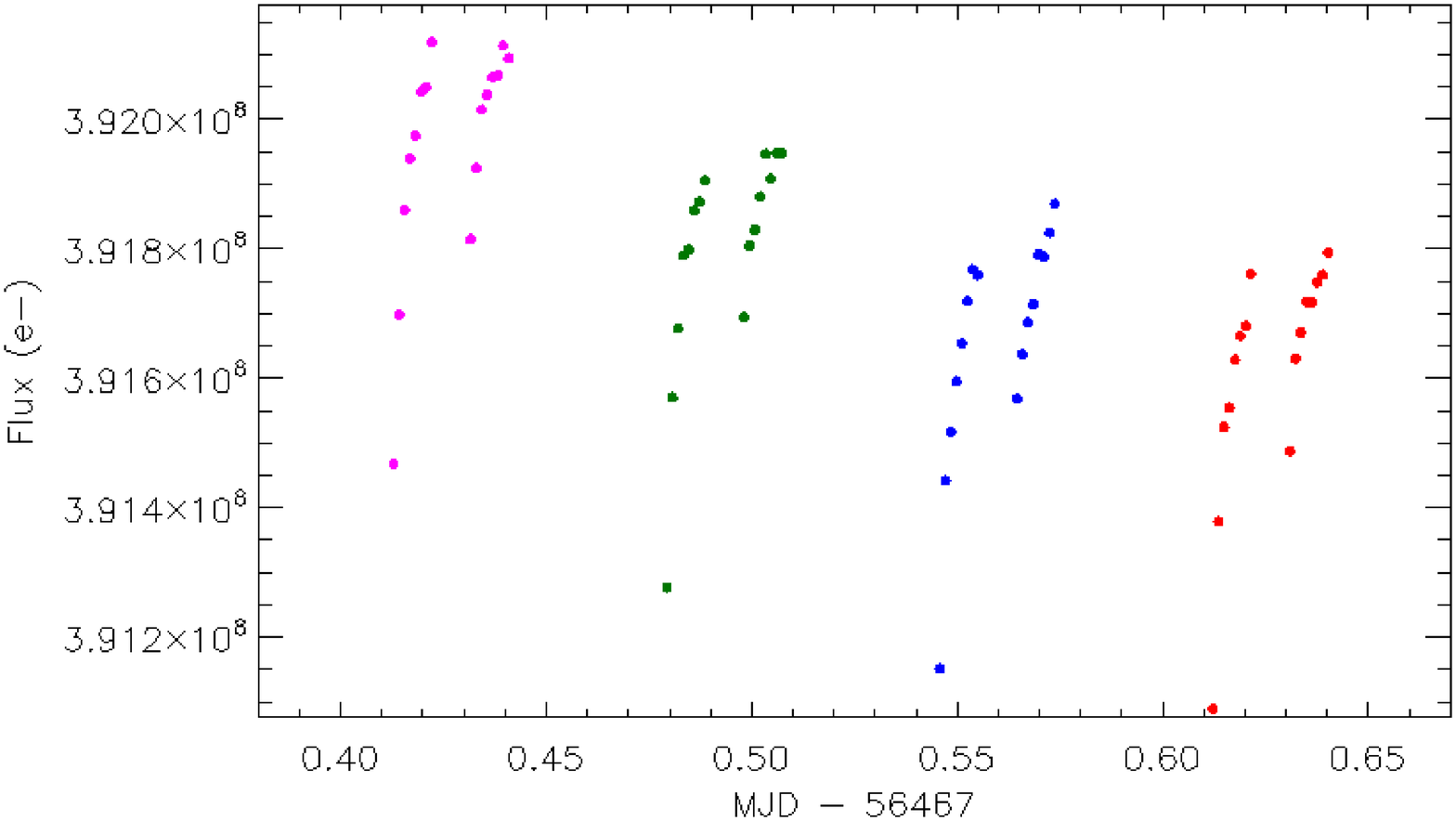}
   \includegraphics[width=4.1cm]{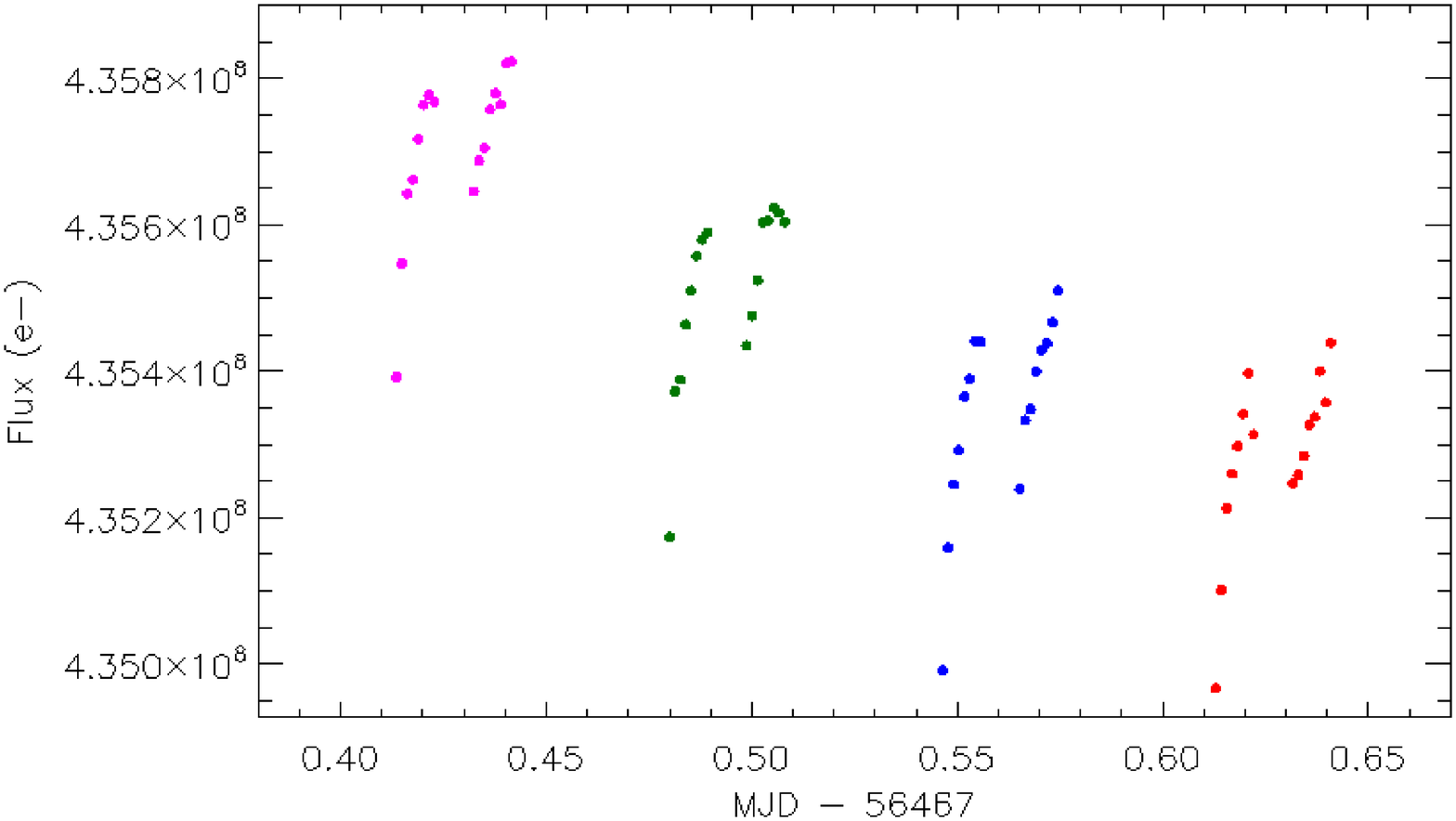}
   \includegraphics[width=4.1cm]{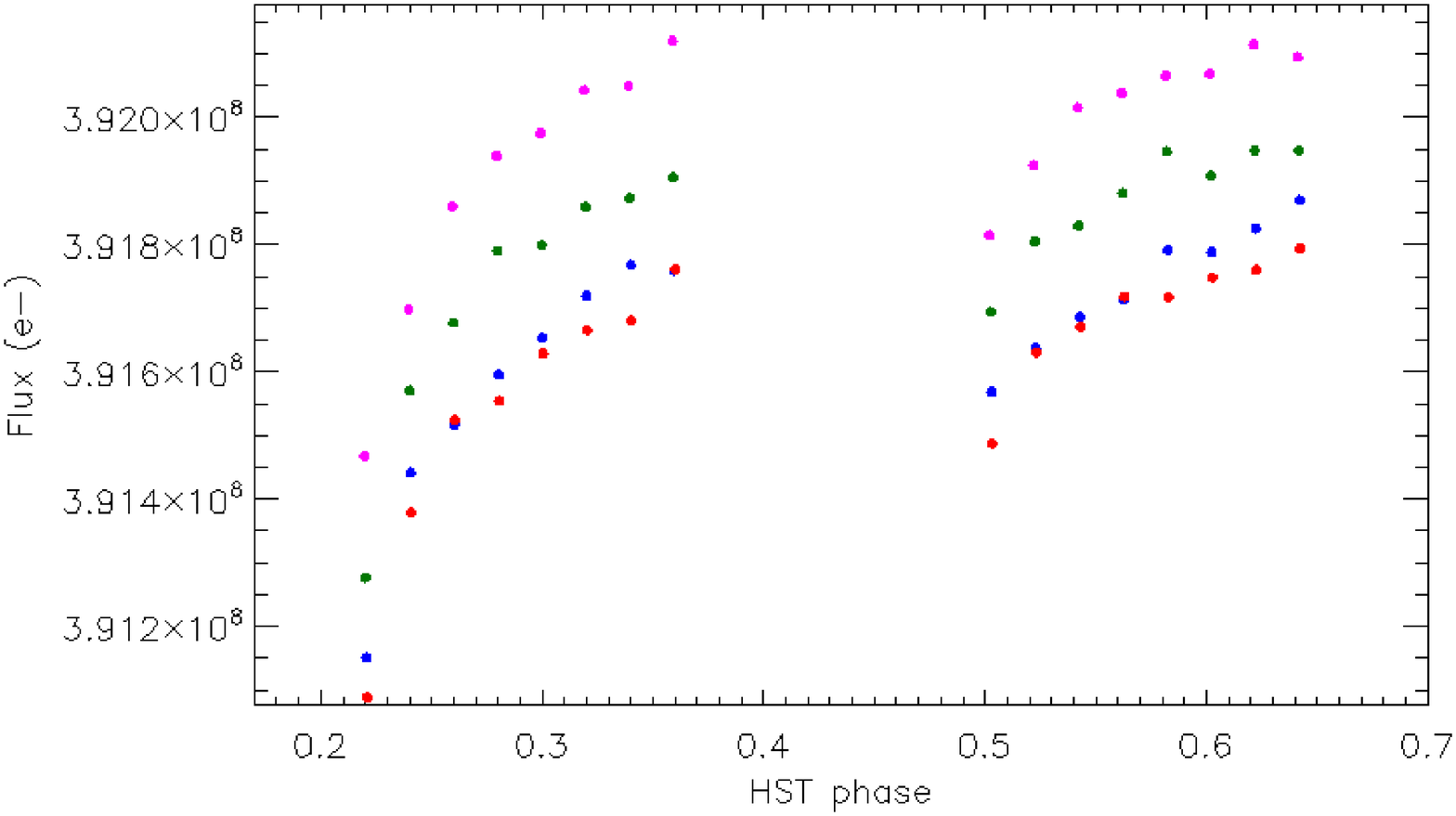}
   \includegraphics[width=4.1cm]{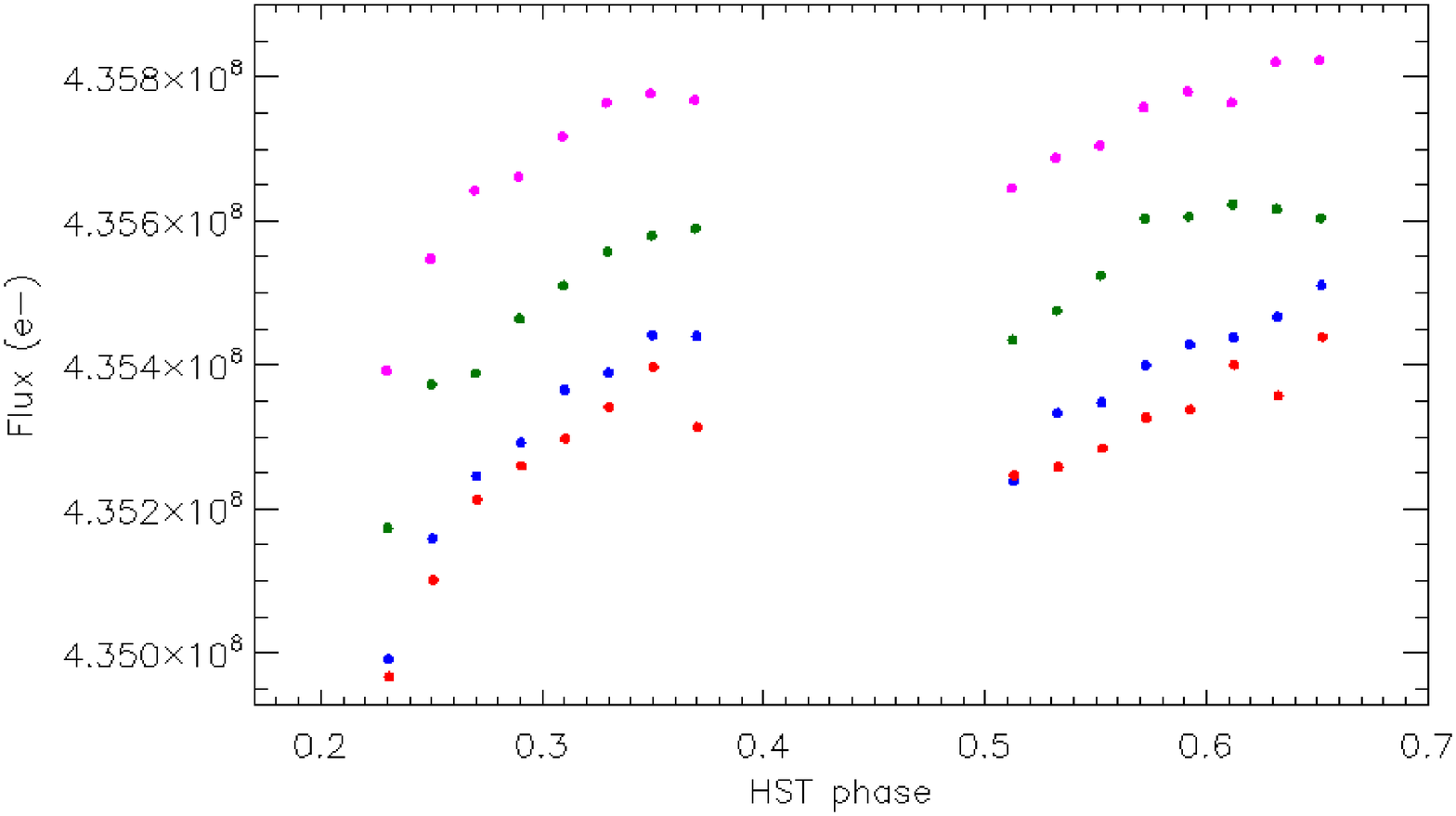}
      \caption{Whitelight curve as a function of time (top) and of \HSTsp orbital phase (bottom) for the forward (left) and reverse (right) scan direction for the 2nd (magenta), 3rd (green), 4th (blue), and 5th (red) orbit. The flux is 11\% greater in the reverse scan direction than in the forward scan direction, increases by \simi0.1\% between each buffer dump, and decreases by \simi0.1\% from the second to the fifth orbit.}
   \label{fig: whitelight all}
\end{figure}

We build a whitelight curve by summing the flux over the stellar region and subtracting the sky background, using the same stellar and sky regions as for the spectral analysis. The fluxes measured at all 7 positions are summed together to recover the flux collected during a full scan. We build a lightcurve for each scan direction and eliminate the first \HSTsp orbit. The resulting lightcurves are reported in Table \ref{tab: whitelight curve}, and are shown in Figure \ref{fig: whitelight all} as a function of time and of \HSTsp orbital phase. From the second to the fifth orbit, observations within each orbit are arranged similarly with respect to the \HSTsp orbital phase. Preparing the observations in a repeatable manner is helpful to correct for systematic effects such as the ``hook", as described next. A good practice is to use a fixed cadence with no interruptions except the Earth occultation, if practical. In our case, a buffer dump interrupts each orbit, with 8 pairs of scans before and 8 pairs after the dump. In these lightcurves, three main features appear. First, the white-light flux is larger by 11\% in the reverse scan direction with respect to the forward scan direction. This is consistent with the longer exposure time in the reverse scan direction, which collects more photons \citep{McCullough2012}. As a result, we analyze each scan direction separately. Second, the flux ramps upward by \simi0.1\% after each buffer dump, which occurs between and in the middle of each orbit. This effect has already been reported for WFC3 and named ``ramp" or ``hook" \citep{Berta2012,Deming2013}. This hook is nearly repeatable from the second to the fifth orbit. Third, the flux decreases by \simi0.1\% from the second to the fifth orbit.

\begin{table*}
\begin{center}
\caption{Whitelight curve}
\label{tab: whitelight curve}
\begin{tabular}{cccrcc}
\tableline
\tableline
EXPSTART MJD   & EXPSTART HJD  & Orbit & Scan & Photo-electrons & Normalized flux \\
56467.413080  &  56467.916605  &  2  &   1  &     391466756  &  0.99927  \\
56467.413739  &  56467.917265  &  2  &  -1  &     435391212  &  0.99985  \\
56467.414399  &  56467.917924  &  2  &   1  &     391697800  &  0.99986  \\
56467.415059  &  56467.918584  &  2  &  -1  &     435546624  &  1.00020  \\
56467.415718  &  56467.919244  &  2  &   1  &     391859136  &  1.00027  \\
56467.416378  &  56467.919904  &  2  &  -1  &     435641776  &  1.00042  \\
56467.417038  &  56467.920563  &  2  &   1  &     391938520  &  1.00047  \\
56467.417698  &  56467.921223  &  2  &  -1  &     435661104  &  1.00047  \\
56467.418357  &  56467.921883  &  2  &   1  &     391973876  &  1.00056  \\
56467.419017  &  56467.922543  &  2  &  -1  &     435716676  &  1.00059  \\
\tableline
\end{tabular}
\end{center}
\vspace{-2mm}
Notes.
The printed table is a truncated version of the electronic table, to illustrate the format.
Columns, left to right, are modified Julian date of the start of the exposure,
the associated heliocentric Julian date, the \HSTsp orbit in the visit, the scan direction
(1 = forward; -1 = reverse), the total number of photoelectrons from HD 189733, 
and the associated normalized flux before detrending. 
\end{table*}

\begin{figure}[htbp]
   \centering
    \includegraphics[width=4.1cm]{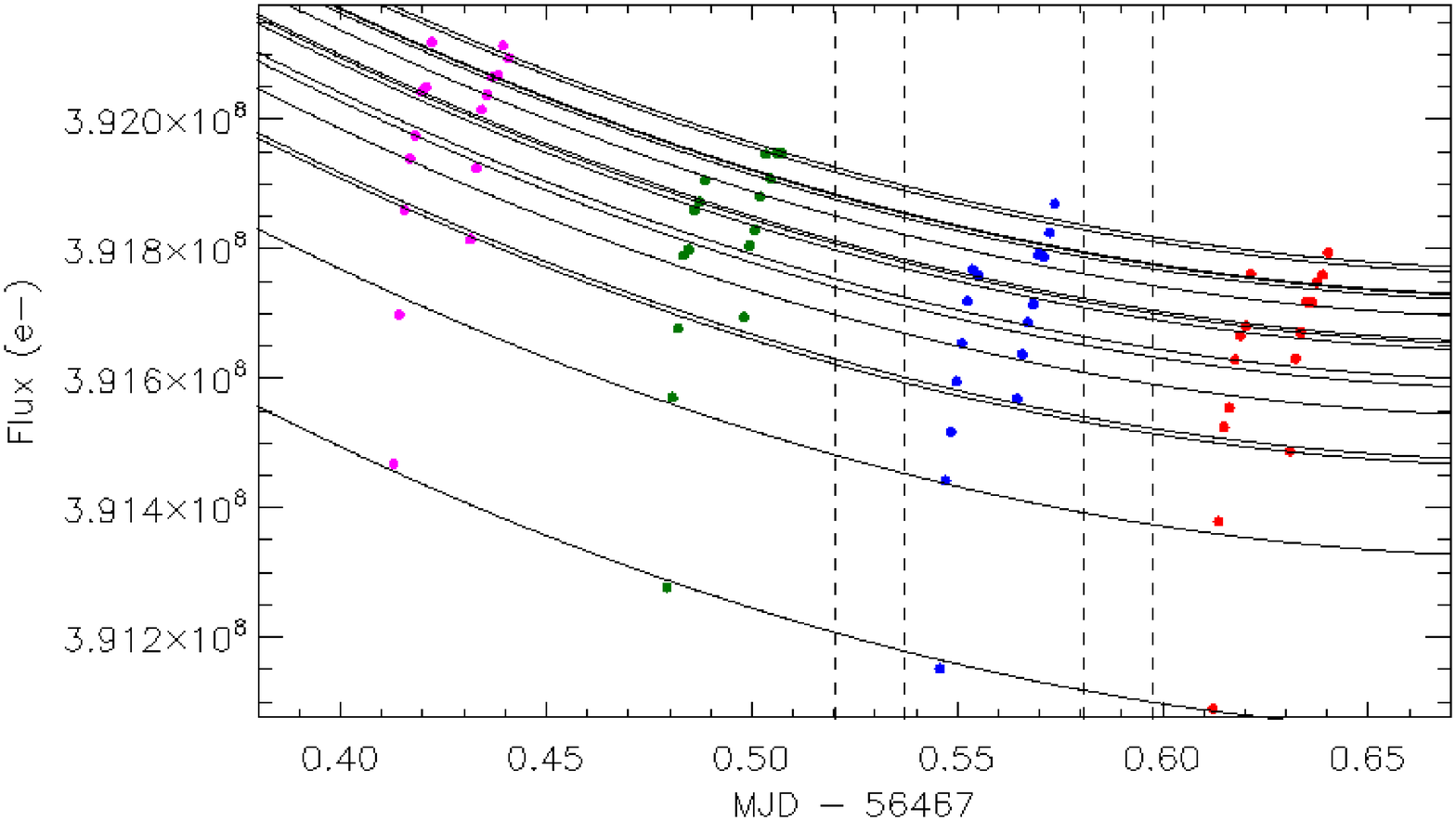}
    \includegraphics[width=4.1cm]{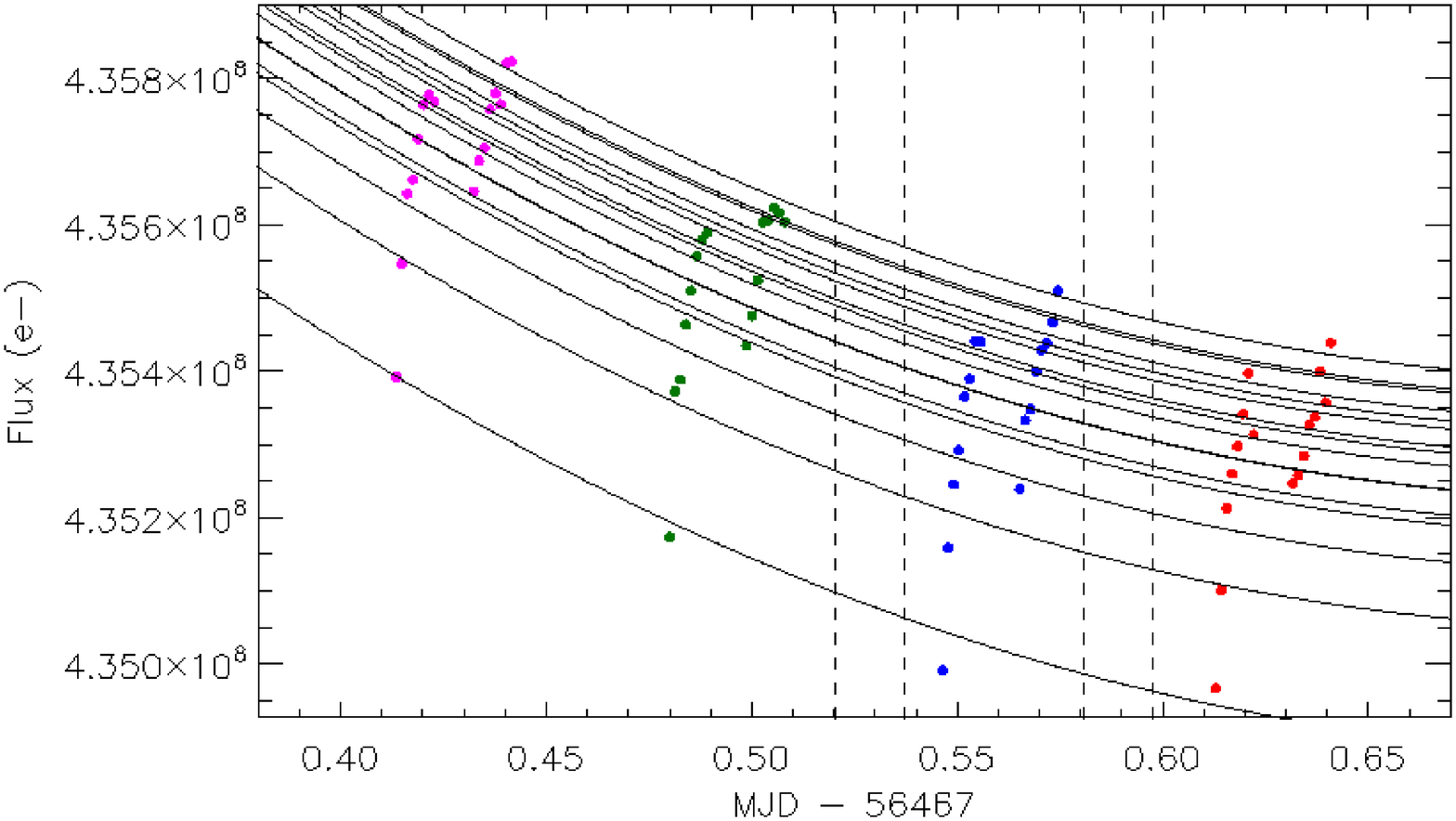}
    \includegraphics[width=4.1cm]{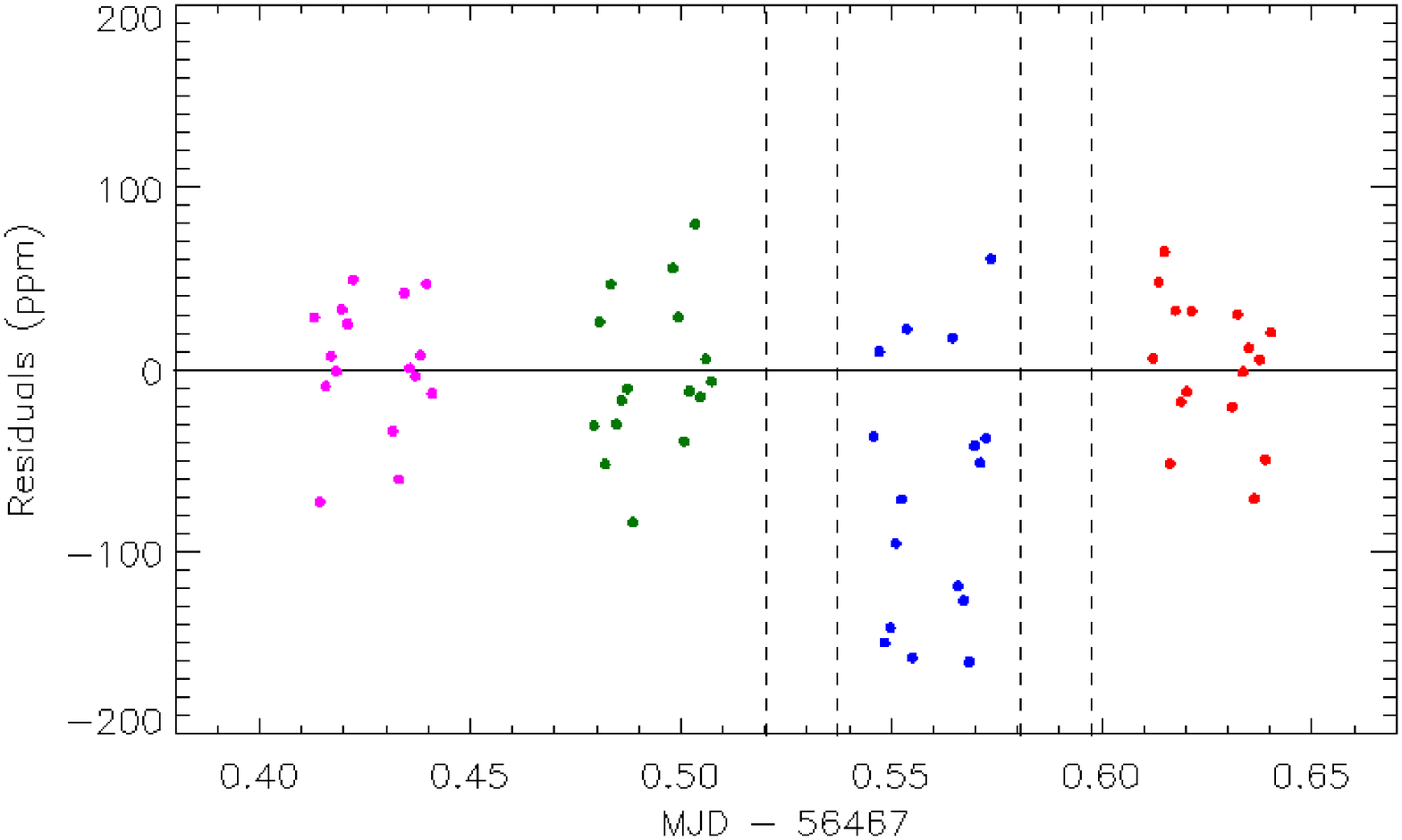}
    \includegraphics[width=4.1cm]{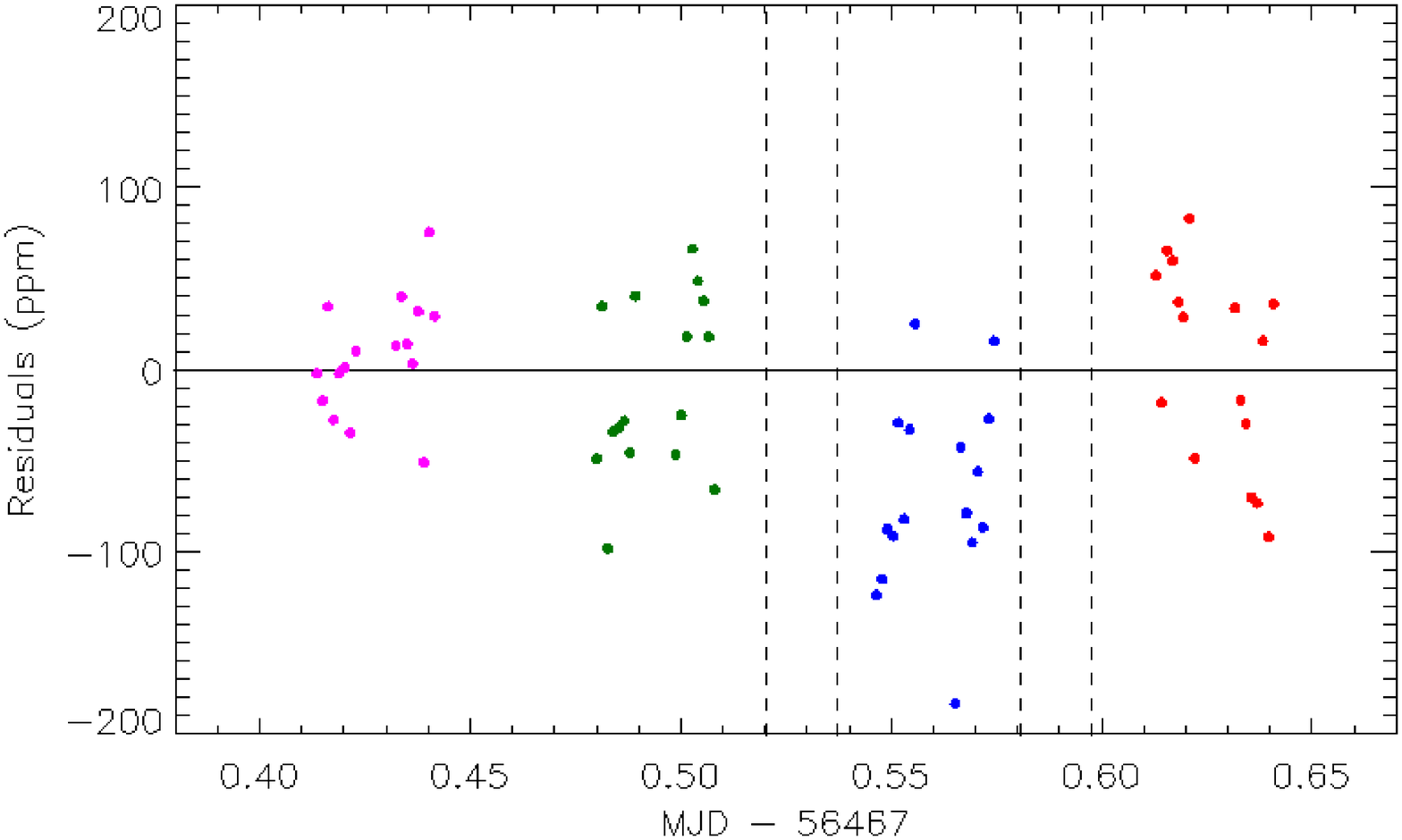}
      \caption{Top: white-light curve as a function of time for the forward (left) and reverse (right) scan direction for the 2nd (magenta), 3rd (green), 4th (blue), and 5th (red) orbit. The vertical dashed lines indicate the 1st, 2nd, 3rd, and 4th contact of the planetary eclipse, from left to right. We model the decrease in time by fitting the same function, here a second order polynomial, to each set of points taken at the same \HSTsp orbital phase using the out-of-eclipse orbits, with the flux offset as a free parameter to account for the hook (plain lines). Bottom: residuals after fitting a second order polynomial.}
   \label{fig: whitelight time poly}
\end{figure}

To derive the eclipse depth, we model the flux's decrease in time first using a second order polynomial. We do not directly model the hook. Instead, we fit the same polynomial function to all the out-of-eclipse points, with the zero flux level as a free parameter for each set of points located at the same position in the orbital sequence (set 1: 1st point of orbits 2, 3, and 5; set 2: 2nd point of orbits 2, 3, and 5; etc..., see Figure \ref{fig: whitelight time poly}). The in-eclipse points are excluded from the fit (orbit 4). This fit is then subtracted to the data points. The residuals for both scan directions are normalized separately and put together. These residuals are close to the Poisson noise. We calculate the eclipse depth $\delta_p$ as the mean of the in-eclipse residuals, and its 1-$\sigma$ uncertainty as:

\begin{equation}
\sigma= \sqrt{\frac{\sigma_{in}^2}{N_{in}}+\frac{\sigma_{out}^2}{N_{out}}}
\end{equation}
where $\sigma_{in}$, $\sigma_{out}$ is the standard deviation of the in-eclipse and out-of-eclipse residuals, and $N_{in}$, $N_{out}$ is the number of in-eclipse and out-of-eclipse points ($N_{in}=32$, $N_{out}=96$). We find $\delta_p=68 \pm 12$ ppm.

Because the choice of a second order polynomial is arbitrary, we perform the same analysis using other functions: an exponential function from the second to the fifth orbit, a linear function from the second to the fifth orbit, and a linear function from the third to the fifth orbit (orbits surrounding the eclipse). Inspection of residuals show that the linear function from the second to the fifth orbit is clearly a poor fit; we exclude this function\footnote{A similar decrease is also present in the transit data \citep{McCullough2014} and in that case is well represented by a linear fit from the second to the fifth orbit.}. For the exponential fit, and the linear fit from the third to the fifth orbit, we find an eclipse depth $\delta_e=79 \pm 12$ ppm and $\delta_l=122 \pm 12$ ppm respectively. In their analysis of WFC3 data of WASP-12b obtained in staring mode, \citet{Stevenson2014} also noticed that different ramp models yield different values for the transit depth (note that in our approach the ramp model is implicit). 

For each model, we estimate the error on the eclipse depth caused by stellar variability, thought to originate mainly from star spots. We model this variability by a sinusoidal function of period 11.95 d \citep{Henry2008}, and rescale the amplitude of 1.5\% in the b+y band \citep{Knutson2012} to our wavelength of 1.4 \m m, using a blackbody emission function with a stellar effective temperature of 4980~K and a spot temperature of 4250~K \citep{Pont2008}. We find an amplitude of 0.8\% at 1.4 \m m. This stellar variation is added to the data at all possible phases of the sinusoidal function prior to performing the function fit. The maximum difference on the eclipse depth with or without this variability is negligible for the second order polynomial and the exponential fits (0.1 ppm), and is 5 ppm for the linear function fit from the third to the fifth orbit, with a standard deviation of 1.7 ppm. We quadratically add this standard deviation to the 1-$\sigma$ uncertainty for this latter function (which result in a change of a fraction of ppm). 

The actual function being unknown, we include the full range spanned by the three models in our final estimate of the eclipse depth $\delta$. This yields $\delta=96\pm39$ ppm. We derive an approximate brightness temperature $T_b$ for the planet by:

\begin{equation}
\delta=\frac{R^2_p B(T_b)}{R^2_\star B(T_{eff,\star})}
\end{equation}
where $R_p$ and $R_\star$ are the planet and stellar radii respectively, \modif{and $B(T_b)$ and $B(T_{eff,\star})$ are the spectral radiance of blackbodies of temperature $T_b$ and $T_{eff,\star}$ respectively}. For simplicity, we approximate the stellar spectrum by a blackbody emission of temperature $T_{eff,\star}=4980$~K (a more accurate model is used in Section~\ref{sec: Discussion}). We find a brightness temperature $T_b=1419_{-96}^{+71}$~K in the wavelength range $1.12-1.65$ \m m for the day-side of \hd b.

\section{Discussion}
\label{sec: Discussion}

We model the dayside atmosphere of HD 189733b using the atmospheric modeling and retrieval method of \citet{Madhusudhan2009} and \citet{Madhusudhan2012}. \modi{The emission spectrum is calculated by solving for} 1D line-by-line radiative transfer in a plane parallel atmosphere, with constraints of local thermodynamic equilibrium (LTE), hydrostatic equilibrium and global energy balance. \modi{The model has twelve free parameters including a parametric temperature profile and molecular abundances parameterized as uniform mixing ratios in the atmosphere. The model includes opacity contributions due to all the dominant molecules expected in H$_2$-rich hot Jupiter atmospheres in oxygen-rich as well as carbon-rich regimes \citep[see \textit{e.g.}][]{Madhusudhan2012}, namely, line opacity due to H$_2$O, CO, CO$_2$, CH$_4$, HCN, and C$_2$H$_2$, and H$_2$-H$_2$ collision induced opacity.} As for the temperature structure, the parametric temperature profile can model dayside temperature profiles with and without thermal inversions. Given a dataset, we explore a wide range of inversion and non-inversion models as well as varied molecular mixing ratios in search of regions in parameter space that best match the data. In the present case, we explore models that can simultaneously explain our WFC3 dataset along with previously published {\it Spitzer} observations in six photometric bandpasses \modif{\citep{Charbonneau2008,Agol2010,Knutson2012}}. 

Considering our present WFC3 observations alone (Figure \ref{fig: final spectrum}), we find nominal evidence for water absorption in the WFC3 bandpass ($1.1-1.7 $ \m m). Figure \ref{fig: final spectrum} shows model spectra that provide a good match to the data. In the temperature regime of HD~189733b, as shown by the $P$-$T$ profile in Figure \ref{fig: multi wavelength spectrum}, several spectroscopically strong molecules are expected to be prevalent in the atmosphere for a solar abundance composition, e.g. H$_2$O, CO, CH$_4$, and CO$_2$. However, H$_2$O is the most abundant of the molecules and dominates the absorption in the WFC3 bandpass, followed by a weaker contribution from CH$_4$, CO, and CO$_2$ which are unconstrained by the WFC3 data. Our data can be explained by a double-trough water absorption feature in the emission spectrum of a solar composition model atmosphere. \modi{On the other hand, given the  error bars in our data, a featureless blackbody spectrum also provides a very good fit to the data. The best-fit blackbody spectrum, corresponding to an isothermal atmosphere at $T=1435$~K, has a $\chi^2$ of 24.8 for 26 degrees of freedom. Therefore, our observational uncertainties preclude a robust constraint on the water abundance using our WFC3 data alone.} 

Combining our data with previously published {\it Spitzer} photometric observations \modif{from \citet{Charbonneau2008,Agol2010,Knutson2012}} rules out an isothermal atmosphere or one with a thermal inversion (Figure \ref{fig: multi wavelength spectrum}). \modif{Even the best-fit blackbody spectrum representing an isothermal atmosphere at $T = 1295$~K} is unable to explain all the existing data simultaneously. \modif{Furthermore, a thermal inversion, \textit{i.e.} with temperature increasing with decreasing pressure,} is conclusively ruled out by the data as such a model will predict even higher fluxes in all the Spitzer IRAC bands than observed. \modif{On the other hand, the sum-total of data can be explained by a dayside atmosphere with no thermal inversion and a nearly-solar or sub-solar abundance H$_2$O mixing ratio.} \modif{Two representative best-fitting models with nearly solar-abundance H$_2$O mixing ratios are shown in Figure \ref{fig: multi wavelength spectrum} with their associated pressure-temperature profiles. The corresponding abundances are reported in Table \ref{tab: abundances}, as well as the statistical significance of each model fit. In this table, the BIC is the Bayesian Information Criterion, which takes into account the number of free parameters in evaluating a model fit. It is calculated as follows: BIC $= \chi^2 + k\,ln\,N$, where $k$ is the number of free parameters (12 for the non-isothermal atmosphere models and 1 for the blackbody), and $N$ is the number of data points (34, \textit{i.e.} 28 from WFC3 and 6 from {\it Spitzer}). The comparison of these models to the data in a $\chi^2$ sense or using the BIC shows that the atmosphere models with molecular features provide significantly better fits than the best blackbody model. 

\modif{A variety of non-isothermal atmosphere models with a range of compositions can also fit the data. The two models shown in Figure \ref{fig: multi wavelength spectrum} \modi{and Table \ref{tab: abundances}} are examples chosen because of their chemical compositions which come close to a solar-abundance H$_2$O mixing ratio while still providing good fits to the data; an exactly solar abundance composition would have a H$_2$O mixing ratio between $5\times10^{-4}$ and $10^{-3}$ depending on the temperature. More generally, however, the space of best-fitting solutions prefer manifestly sub-solar H$_2$O abundances\modi{: the 1-$\sigma$ range of H$_2$O abundance derived from the sum-total of data used is between $1.2\times 10^{-9}$ and $2.3\times 10^{-5}$, with a modal value of $7.6\times 10^{-6}$}. Our H$_2$O abundances are generally consistent, at $\sim$2-$\sigma$, with other studies in the past also reporting constraints that are consistent with a sub-solar H$_2$O abundance on the dayside of HD 189733b \citep[e.g.,][]{Madhusudhan2009,Swain2009,Lee2012,Line2012,Line2014} using previous {\it Spitzer} and {\it HST} data, some of which have since been revised. However, our 1-$\sigma$ limits suggest lower H$_2$O abundances than previously suggested. \mod{Note that low altitude, thin, or patchy clouds, and/or haze in the planet's atmosphere could also damp spectral features, as discussed below. Since our models are cloud-free, the molecular abundances derived here should be interpreted as lower limits if obscuring clouds are present.}} 

The lack of a thermal inversion in the dayside atmosphere of HD~189733b is consistent with previous studies, both observational and theoretical. Several studies using past data have suggested the absence of a thermal inversion in HD~189733b \citep[\textit{e.g.}][]{Burrows2008,Grillmair2008,Swain2009,Madhusudhan2009}. Several theoretical and empirical studies have also predicted that its dayside atmosphere would be unlikely to host a strong thermal inversion. With an equilibrium temperature of $\sim 1200$~K, HD~189733b is one of the less irradiated hot Jupiters. Based on the original TiO/VO hypothesis \citep{Hubeny2003,Fortney2008}, cooler hot Jupiters such as HD~189733b would not be hot enough to host gaseous TiO and VO in their upper atmospheres which have been proposed as inversion-causing compounds. Even if alternate inversion-causing compounds were possible, \citet{Knutson2010} suggested that the high chromospheric activity of the host star (HD~189733) might dissociate them in the planetary atmosphere.

\begin{figure*}[htbp]
   \centering
   \includegraphics[width=12cm]{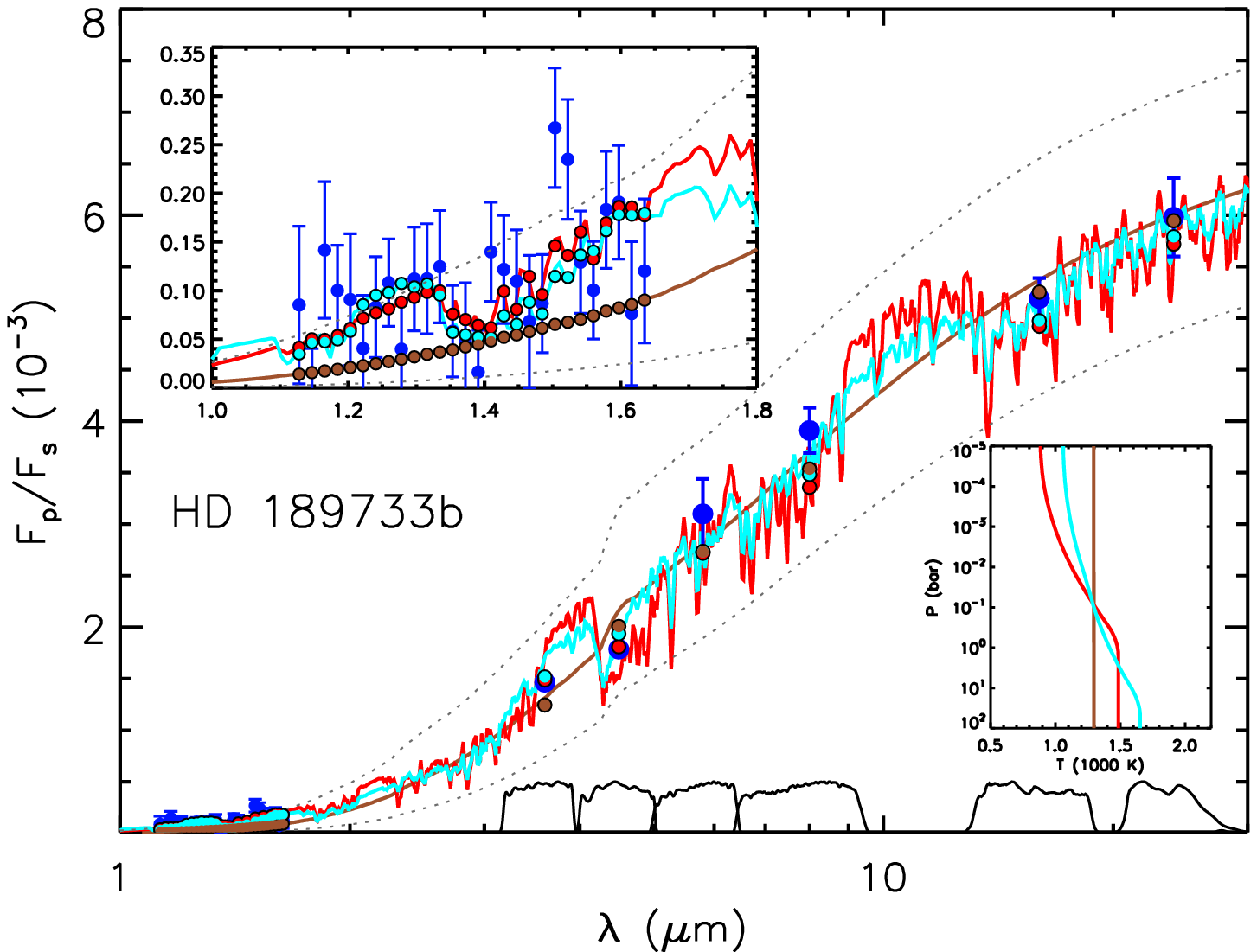}
      \caption{\modif{Observations and model spectra of day-side thermal emission from HD~189733b. The observations are show in blue circles with error bars, and include data obtained using WFC3 in the 1.1-1.7 $\mu$m range (this work) and {\it Spitzer} photometry at longer wavelengths \modif{\citep{Charbonneau2008,Agol2010,Knutson2012}}. Two theoretical model spectra are shown in the \modi{red} and \modi{cyan} curves corresponding to chemical compositions shown in Table~\ref{tab: abundances} \modi{(model 1 and model 2 respectively)} and non-inverted temperature profiles shown in the lower-right inset. The brown curve shows a model blackbody spectrum representing an isothermal atmosphere at 1295~K. The band-pass integrated model points for each model are shown in the same colored circles. The dashed gray lines show blackbody spectra at 1100~K (bottom) and 1500~K (top). The vertical axis is the ratio of planetary to stellar flux. The top left inset zooms in on the WFC3 spectrum, and the bottom right inset shows the atmospheric pressure-temperature profile for the three models.}}
\label{fig: multi wavelength spectrum}
\end{figure*}

\begin{table}
\begin{center}
\caption{Molecular abundances relative to H$_2$ and statistical significance for \modi{two fitting models and a blackbody model, for the combination of WFC3 and Spitzer data.}}
\begin{tabular}{lccc}
\hline
  & \modi{Model 1} & \modi{Model 2} & Best-fit   \\
  &               &              & blackbody \\
\hline
H$_2$O & $1.0\times 10^{-4}$Ê& $1.1\times 10^{-4}$Ê& - \\
CO & $9.4\times 10^{-3}$Ê& $3.4\times 10^{-4}$Ê& - \\
CH$_4$ & $7.9\times 10^{-6}$Ê& $7.9\times 10^{-8}$Ê& -  \\
CO$_2$ & $1.6\times 10^{-6}$Ê& $5.0\times 10^{-6}$Ê& -  \\
\vspace{-2.5mm} \\
$\chi^2$ & 31.7  & 51.2  & 122.1  \\
d.f.   &  21 & 21 & 32 \\
BIC & 74.0 & 93.5 & 125.6 \\
\hline
\vspace{-5mm}
\label{tab: abundances}
\end{tabular}
\end{center}
Notes. C$_2$H$_2$ and HCN are also present, but are negligible. $\chi^2$ are non-reduced $\chi^2$. \modi{d.f. indicates the number of degrees of freedom.}
\end{table}

The ensemble of data obtained on \hd b \modif{in transmission and emission} from \HSTsp and \textit{Spitzer} is yet to be understood completely. \modif{\citet{Pont2013} interpreted the combined transmission spectrum} by an atmosphere dominated by dust and clouds. They also noted that several puzzling features arising from the interpretation of the \textit{Spitzer} phase curves with a clear solar composition atmosphere model \citep{Knutson2012} could be explained by clouds \citep[see Table 8 of][]{Pont2013}. \modif{However, \citet{McCullough2014} recently reported a robust detection of water vapor absorption in the transmission spectrum of the planet, which rules out a featureless transmission spectrum, as well as an atmosphere dominated by opaque a clouds at pressures $P > 0.1$ bar at the terminator region. Instead of clouds, \citet{McCullough2014} interpret the multi-wavelengths transmission spectrum by a solar abundance composition clear atmosphere model with an additional contribution of unocculted star spots explaining the rise in the visible and UV.} 

\modif{In emission, a higher geometric albedo at short visible wavelengths ($\lambda < 0.45$ \m m) than at longer wavelengths ($\lambda > 0.45$ \m m) was reported by \citep{Evans2013} who suggested an ``intermediate cloud scenario" for the day-side of the planet in which clouds are present and become optically thick at pressures corresponding to the Na absorption wings. Recent theoretical studies have also suggested the possibility of clouds in the dayside atmospheres of hot Jupiters \citep[e.g.][]{Heng2013,Parmentier2013}. However, as shown by \citet{Barstow2014}, the reported albedo spectrum is insufficient to conclusively constrain the presence of clouds in the dayside atmosphere of HD 189733b, as a cloud free atmosphere with Rayleigh scattering due to H$_2$ and non-solar Na abundances could explain the data equally well. Our detection of molecular features in the WFC3 bandpass places additional constraints on the possibility of clouds on the dayside. With thick clouds on the dayside, we would expect a blackbody spectrum to fit the observed infrared emission spectrum better than a model spectrum with molecular features in the observed bandpasses; our retrieved model solutions favor the opposite. Interestingly, a new analysis of the {\it Spitzer} IRS data including data not included in the original spectrum from \citet{Grillmair2008} is also inconsistent with a blackbody spectrum \citep{Todorov2014}. In principle, alternate scenarios might be able to explain both the reported albedo spectrum and the molecular features we are observing, e.g., a layer of clouds at low altitudes producing the former and a clear atmosphere at higher altitudes producing the latter, or Rayleigh scattering due to H$_2$ with non-solar Na abundances, which future retrieval studies using clouds models could investigate \citep[e.g.][]{Barstow2014,Lee2014}.}

The transit and eclipse spectra of \hd b obtained with the high precision of WFC3 in spatial scanning mode yield new constraints both on the limb and day-side regions of its atmosphere. These results support a general picture of an atmosphere dominated by water vapor in the near infrared, albeit with a lower H$_2$O abundance than for a solar composition atmosphere. \mod{If clouds are present, these abundances would be degenerate with the amount of cloudiness; however, the presence and composition of such clouds are not constrained observationally \citep[see][for a discussion on this low H$_2$O abundance]{Madhusudhan2014}.} These results point towards the need for including the effects of clouds in hot Jupiter atmosphere models, although still poorly understood due to their extreme complexity. Finally, these transit and eclipse spectra provide keys to further interpretation using full 3-D atmosphere models such as developed by \eg \citet{Showman2009}. Extensive characterization of other hot gaseous giant planets would also improve our understanding of their atmospheres. The TESS mission \citep{Ricker2013} should detect new hot Jupiters and hot Neptunes around bright stars, providing ideal targets for such characterization.   


\section{Summary}
\label{sec: Summary}

We observed \hd b during a planetary eclipse with \HSTsp WFC3 and extracted the emission spectrum of the planet in the wavelength range $1.1-1.7 \mu$m. Using a straightforward data reduction method of the spatially scanned spectra, the derived spectrum is Poisson noise limited. A white-light analysis including a correction for WFC3 instrumental systematic effects yields the absolute eclipse depth in this wavelength range. \modif{The resulting spectrum shows \modi{marginal evidence for water vapor absorption, but can also be well explained by a blackbody spectrum. However, the combination of our WFC3 data} with previous data from \textit{Spitzer} reinforces a general picture of a day-side atmosphere in thermochemical equilibrium with no thermal inversion, and dominated by water vapor features in the near infrared.

\acknowledgments

The authors gratefully acknowledge everyone who has contributed to the {\it Hubble Space Telescope} and the WFC3, and particularly those responsible for implementing the spatial scanning, which was critical to these observations. We thank in particular John MacKenty and Merle Reinhart. We acknowledge conversations with Suzanne Hawley, Leslie Hebb, Veselin Kostov, Rachel Osten, and Neill Reid.
This research used NASA's Astrophysics Data System Bibliographic Services, the SIMBAD database operated at CDS, Strasbourg, France, and was funded in part by \HSTsp grant GO-12881 and Origins of Solar Systems grant NNX10AG30G.


\bibliography{biblio}

\end{document}